# Does MoSE cope with inland tsunamis hazard?

**Giuliano Francesco Panza**[a,b,c,d], **Davide Bisignano**[a], **Fabio Romanelli**[a,b]

[a] Department of Mathematics and Geosciences, University of Trieste, Via Weiss4, 34127, Trieste, Italy.

[b] The Abdus Salam International Centre for Theoretical Physics, Strada Costiera 11, 34014 Trieste, Italy.

[c] Institute of Geophysics, China Earthquake Administration, Minzudaxuenanlu 5, Haidian District, 100081 Beijing, China

[d] International Seismic Safety Organization (ISSO) - www.issoquake.org

**Extended abstract**.

Tsunami hazard from inland earthquakes.
Modeling hazard scenarios has the main purpose to assess the maximum threat expected from a studied phenomenon in a certain area and to give specific information to the local authorities in order to prevent and mitigate serious negative consequences for the population, the infrastructures and the environment. To build, for a specific coastal area, scenario-based tsunami hazard maps it is necessary to characterize the seismic sources (or other tsunamigenic events) and to select the earthquake scenarios that can drive the hazard. By means of modeling it is then possible to calculate the maximum amplitude of the vertical displacement of the water particles on the sea surface and its travel time - the most relevant properties of the tsunami wave - always recorded in the chronicles and therefore in catalogues, thus useful for the validation of the modeling. The horizontal displacement field is calculated too, and, on average, it exceeds the vertical one by an order of magnitude approximately (this accounts for the great inundating power of tsunami waves with respect to wind driven ones).

Due to the distances and the magnitude of the considered sources in earlier studies (e.g. Paulatto et al., 2007, Tiberti et al., 2009) it is known that the tsunami hazard from offshore sources for the city of Venice is relatively low: the expected maximum heights are less than about 50 cm. These results are confirmed also considering the possible tsunamigenic sources – source codes ITCS100 and ITCS101 - described in the updated European Database of Seismogenic Faults (SHARE, 2014). On account of the theoretical results of Yanovskaya et al. (2003) we consider here a potential source, located inland very near to Venice and we evaluate how a tsunami wave generated from this source can affect the MoSE gates if they are standing up (closed) during the tsunami event. From our simulation we get both peaks and troughs as first arrivals: the behavior of the barriers in these two situations could be a very important design matter.

The considered seismic source is mentioned, with some reserve, by Gorshkov et al. (2013) who use morphostructural zonation and pattern recognition techniques to identify, in the Po Plain, areas prone to an earthquake with M≥5.0 (M5+). Therefore, on account of the precautionary principle, the earthquakes source area that should be considered for the study of the tsunami hazard in Venice is the node 80, conventionally defined as a 25 km radius circle, centered at coordinates 45.49° N, 12.29° E, and indicated by a red arrow in fig.1. In fact, historical information does not allow the exclusion of this node as the location area of the 4 historical epicenters listed in the UCI0912 catalog, all with M > 5 (see tab. 1).

| Date | Latitude | Longitude | Magnitude | Catalog |
|---|---|---|---|---|
| 10-21-1093 | 45.42° | 12.37° | 5.2 | absent in CPTI11 |
| 7-11-1282 | 45.42° | 12.33° | 5.2 | absent in NTT411, and CPTI11 |
| 10-8-1410 | 45.42° | 12.37° | 5.2 | absent in NTT411, and CPTI11 |
| 1-1-1429 | 45.47° | 12.33° | 5.2 | absent in NTT411, and CPTI11 |

Table 1. Venice earthquakes with M5+ from the UCI0912 catalog.

Looking at fig.1 (from Gorshkov et al., 2013) we see that, differently from node 80 for which further studies are needed, nodes 75 and 87 (indicated by blue arrows) are recognized as prone to M5+ earthquake. These two nodes do not represent effective tsunami sources for the Venice lagoon, as it can be proven by modeling

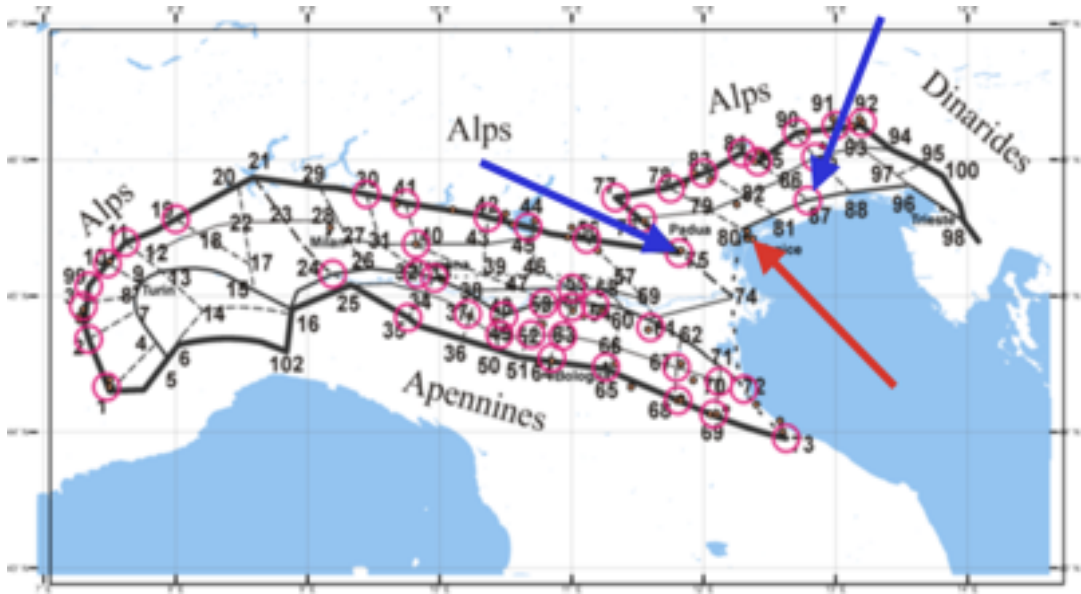

Fig.1 Morphostructural zoning map of the Po plain and earthquakes M5+ from 1093 to 2012. Lines depict lineaments of different rank. Dots mark epicenters M5+ from the UCI0912 catalog. Red circles show recognized nodes prone to M5+ earthquakes by Gorshkov et al. (2013).

    Nevertheless from figs. 2 and 3, it is evident that the structure of the lithosphere underlying these 3 nodes (75, 80 and 87) presents some common features: the depth of the 3.3 g/cm$^3$ isoline and the depth of LID bottom (from Brandmayr, 2013). In the same figures are shown the 3 aforementioned nodes and the epicenter with focal mechanism of the strongest events (M≥4.9) 2012 Emilia Earthquake (from Brandmayr et al., 2013). Following Brandmayr (2013) we can relate the depth of 3.3g/cm$^3$ isoline to the thickness of the LID, as it is evident also comparing figures 3 and 4. More precisely low depth of this isoline corresponds to a shallow asthenosphere-lithosphere boundary.

    In these 2 figures the green dots represent the epicenters of 2012 Emilia Earthquake with associated focal mechanisms, and in particular the 2 easternmost are those of the two large shocks of the 20[th] and the 29[th] of may 2012. From figs. 2 and 3 it is evident that these two epicenters and the node 80 sit on similar lithosphere. These similarities just described, on account of the precautionary principle, makes it worth to consider node 80 as the area of possible scenario earthquakes to assess tsunamis hazard in Venice, from inland earthquakes.

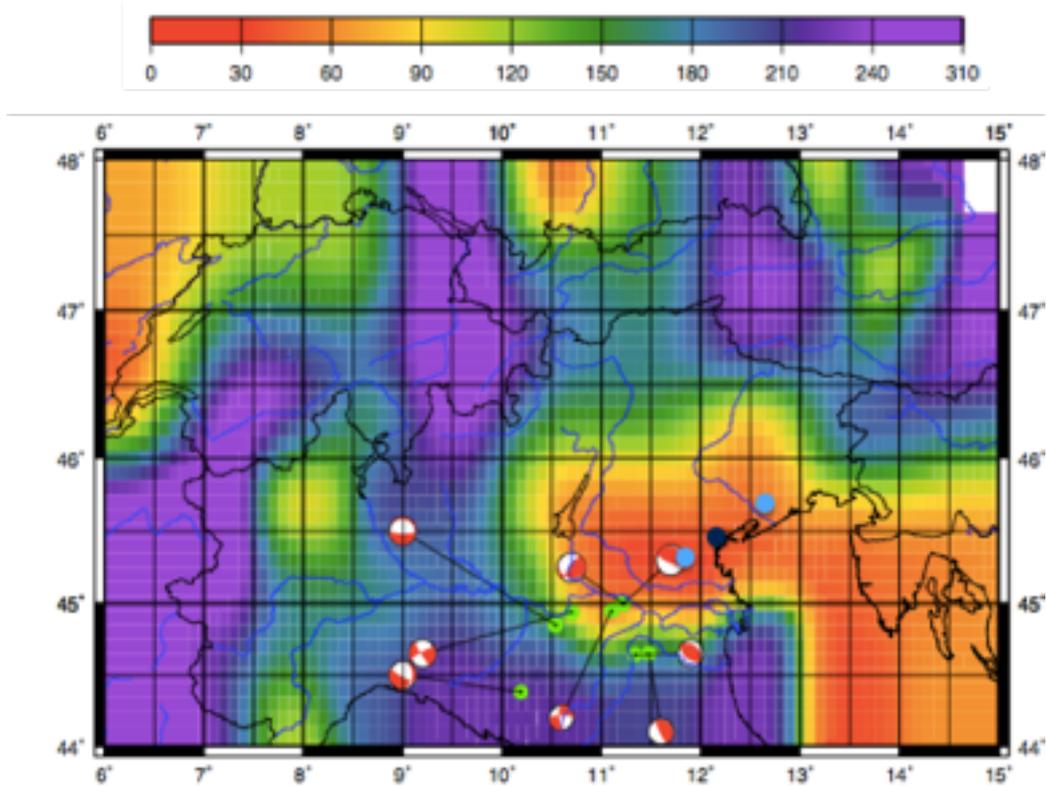

Fig. 2. Depth of 3.3 g/cm3 isosurface determined by density model (Brandmayr, 2013), epicenters (green circles) of Emilia Earthquake with their focal mechanism (Brandmayr et al., 2013) and nodes 80 (dark blue circle), 75 and 87 (light blue circles).

The tsunami wave from this source, located inland, has been simulated using the Green Function approach developed by Yanovskaya et al. (2003). The efficiency of this analytical method allowed us to perform many parametric tests in order to get a general idea of the expected hazard. In detail we have studied the generated waves varying the magnitude, the depth of the source, the dip angle and the distances from the source to the coast and from the coast to the receiver.

For the vertical component, the maximum peak values, among all performed parametric tests, are 0.5 cm and 4 cm for M = 5.5 and M = 6, respectively. For the horizontal component the corresponding peak values are about 5 cm and 40 m, respectively.

For M = 6.5 the maximum vertical peak value is about 25 cm and it has been obtained considering the dip angle of 60°, the depth source of 10km, the distance from the source to the coastline of 3 km and the distance from the coastline to the receiver of 10 km. In correspondence of these parameters, the peak value for the horizontal component is about 2.5 m

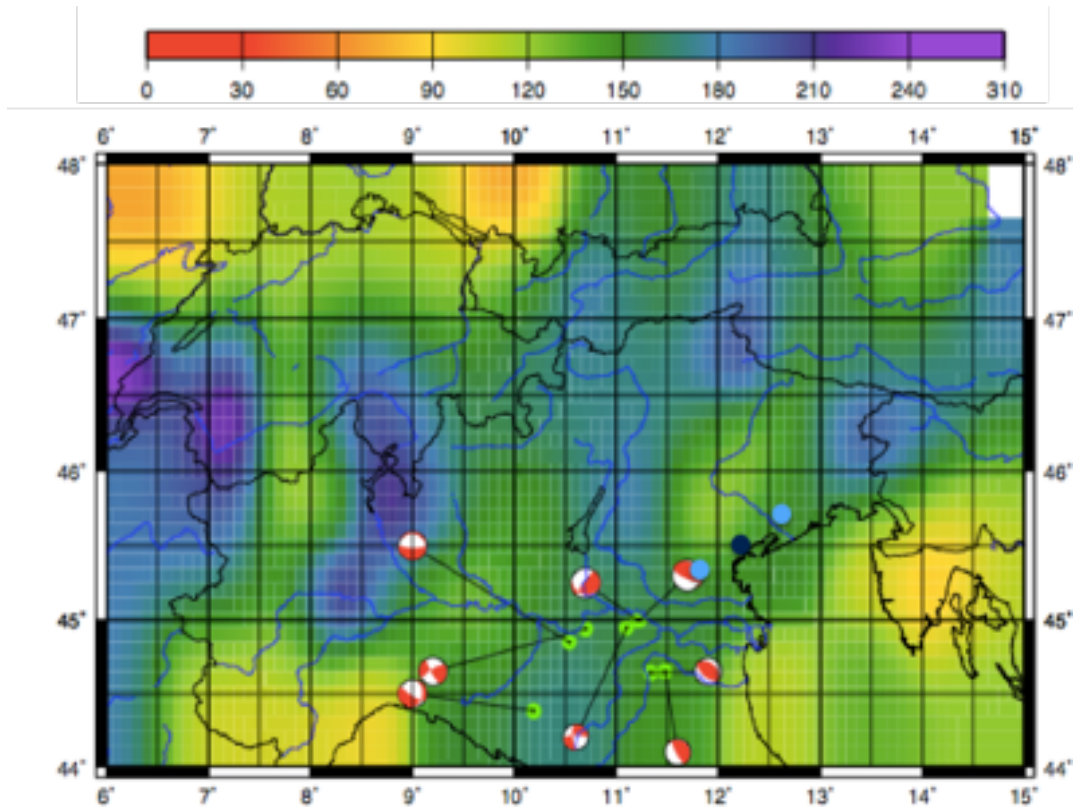

Fig. 3. Depth (km) of the LID bottom determined by Vs (Brandmayr, 2013), epicenters (green circles) of Emilia Earthquake with their focal mechanism (Brandmayr et al., 2013) and nodes 80 (dark blue circle), 75 and 87 (light blue circles).

A particular attention should be posed to this case because, as we'll show later in detail, converting to macroseismic intensity (MCS) the peaks of soil motion computed using this magnitude we obtain a value of VIII that is the reference historical macroseimic degree in the mainland north of Venice.

In tsunami hazard computation the one usually reported on maps, is the vertical component of motion, nevertheless, as we have shown, the horizontal displacement field, on average, exceeds the vertical one by an order of magnitude, approximately, and this fact can represents a crucial aspect in the evaluation of the tsunami waves interaction with MoSE gates.

To be more conservative, and to maximize the precautionary principle, we have computed tsunami waves considering a source with a value of M=7, which, very likely, exceeds the maximum credible earthquakes (MCE) for the study area. In this case the maximum obtained vertical wave peak reaches 70 cm, while the horizontal is about 7 m.

In tab. 2 are listed the vertical peak values for different dip angles and coastline-receiver distances with the strike-receiver angle fixed at 90° and considering a dip-slip mechanism.

|        | 3 km   | 5 km   | 10 km  | 15 km  |
|--------|--------|--------|--------|--------|
| 0°     | 40 cm  | 50 cm  | 50 cm  | 25 cm  |
| 15°    | 25 cm  | 20 cm  | 30 cm  | 25 cm  |
| 30°    | 50 cm  | 40 cm  | 10 cm  | 15 cm  |
| 45°    | 70 cm  | 50 cm  | 20 cm  | 5 cm   |
| 60°    | 70 cm  | 70 cm  | 35 cm  | 8 cm   |
| 75°    | 40 cm  | 60 cm  | 50 cm  | 18 cm  |
| 90°    | 20 cm  | 30 cm  | 35 cm  | 25 cm  |

Table 2. Parametric test on dip angle and source-coastline distance with receiver placed at 10 km from the coastline (corresponding to MoSE barriers position).

In fig.4 the complete signals of the third column of Tab. 2 show that the first arrivals can be a peak or a trough depending on the dip angle. In fig.5 4 signals are plotted with different (from 10 km to 25 km) receiver distance but maintaining fixed the source-coastline distance (10 km) and the dip angle (60°), while in fig.6 we show the maximum tsunami heights distribution for the entire lagoon region using a dip angle of 60° and a source-coastline distance of 10 km.

Figure 5 demonstrates that in case of inland source the wave is outgoing. Moreover considering a liquid layer of 20 m we have, as expected, a tsunami phase velocity of about 14 m/s and so a time interval of about 6 minutes between the peaks at the receivers, equidistant at 5 km.

The results shown in these last two figures together pose relevant questions about the design of MoSE system. In fact the problem can be double, depending on the characteristic, peak or trough, of the first arrival.

In the first case the question can be: are the MoSE gates able to sustain the pressure of a wave hitting them in the opposite direction compared to the one of the sea tide?

On the other hand, when the first arrival is a trough, the relevant question is: considering the fact that the MoSE gates have been designed to stand up also thanks to the Archimede's principle, see fig. 7c, what can happen if the water level decreases significantly due to the tsunami?

Moreover the width of Venice lagoon is about 10 km so we can argue that a tsunami wave generated by an inland source can cross it in less than 15 minutes, which is the time needed to reopen MoSE gates eventually standing up during the tsunami event.

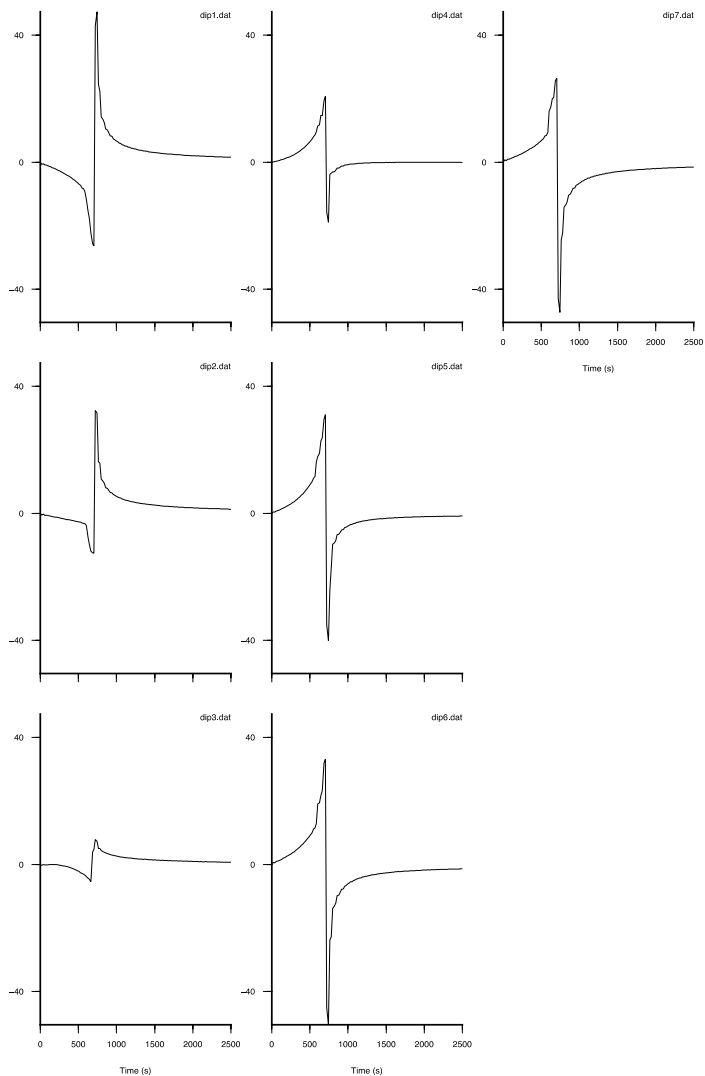

Fig. 4. Parametric test on dip angle (column 3 of Tab.2).

Land ground motion.

Using the web application "WebXeris" developed by Franco Vaccari and eXact lab s.r.l. (http://www.exact-lab.it) and based on the modal summation for seismic wave, the soil motion due to the selected sources has been computed. In particular, the ground displacements and velocities have been computed using 3 different strike-receiver angles (90°, 45° and 0°) in the magnitude range from 5.5 to 7

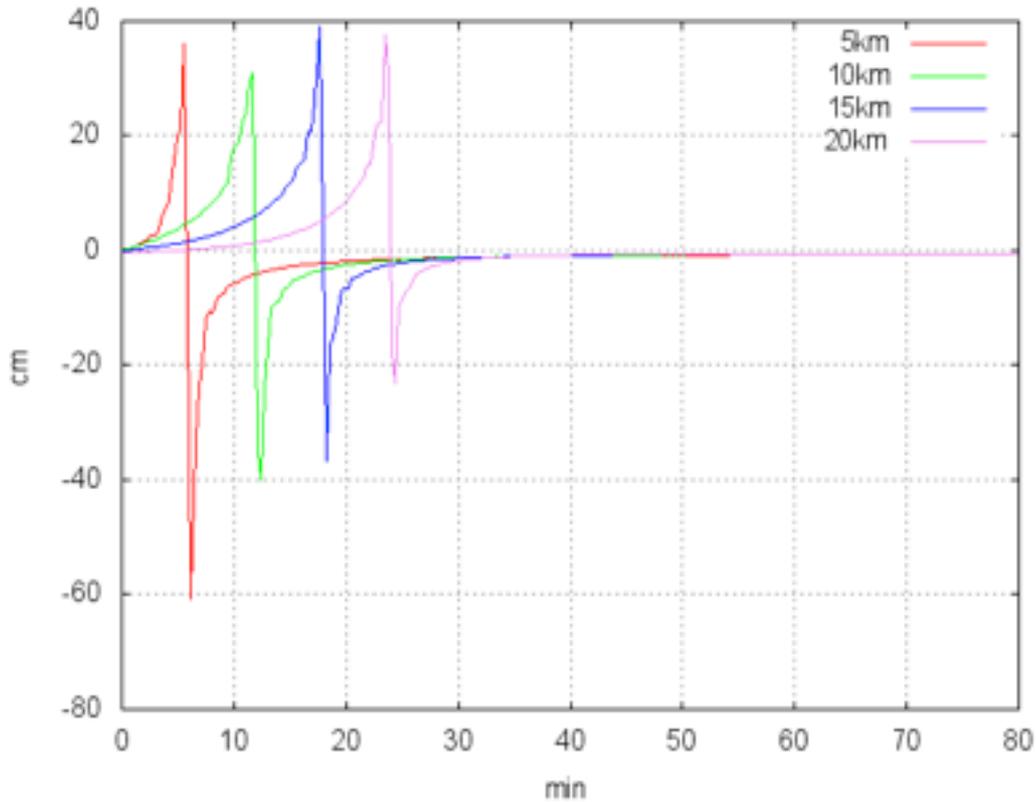

Fig 5. Parametric test on coastline-receiver distance (with M = 7.0, dip angle = 60° and source-coastline distance fixed to 10 km).

 

The results of our computation have been validated with historical information, converting the obtained peak values to macroseismic intensity scale (MCS) degrees.

This convertion has been based on the works of Magrin (personal communication), Magrin (2013), Zuccolo et al. (2011), Panza et al. (2001), Decanini et al. (1995)

The results obtained using a strike-receiver angle of 90° and M=6.5 show that (tab. 3) there is a correspondence with I(MCS)= VIII, that is the observed intensity degree for the mainland north of Venice.

The following tables show, for the three selected strike-receiver angles, the peak velocity values for the considered magnitude range and their conversion into MCS scale.

In table 3 are shown the results for a strike-receiver angle of 90°, in table 4 are shown the results for a strike-receiver angle of 45° while in table 5 there are those obtained for a strike-receiver angle of 0°are those obtained for a strike-receiver angle of 0.

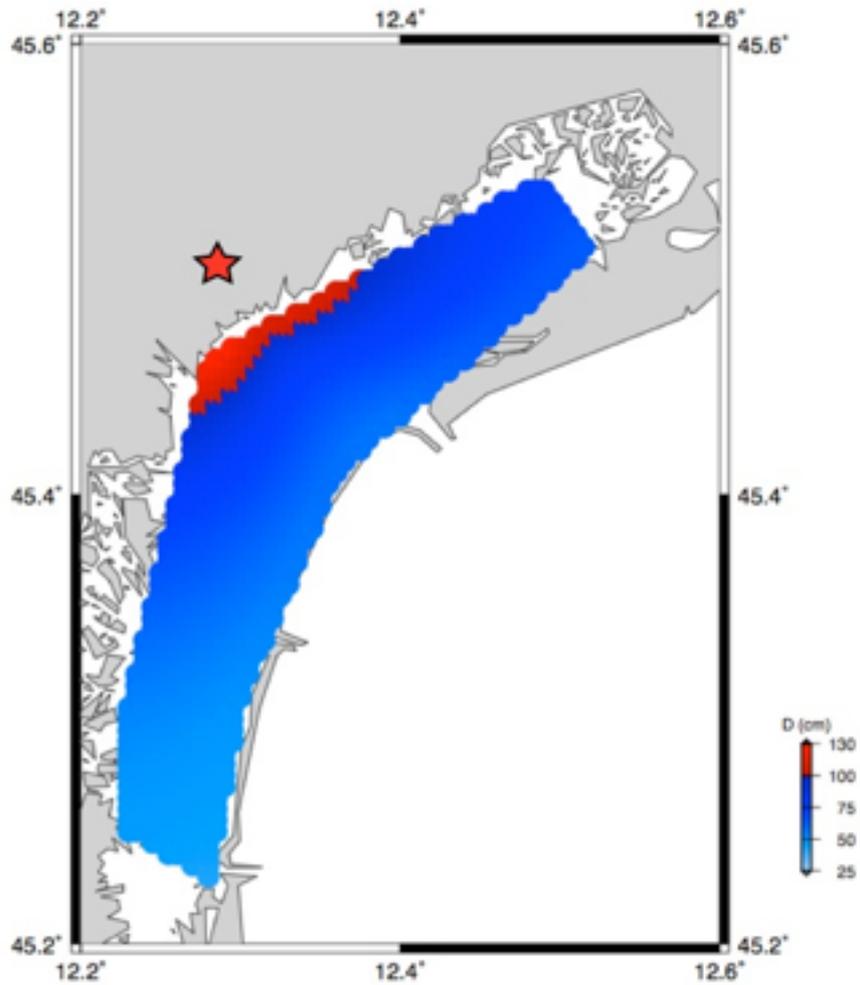

Fig 6. Distribution of the maximum tsunami heights in the Venice Lagoon for the considered source indicated by red star (with M = 7.0, dip angle = 60° and source-coastline distance fixed to 5 km).

Parametric test with Strike-receiver angle = 90°

| Magnitude | 5.5 | 6 | 6.5 | 7 |
|---|---|---|---|---|
| Peak Velocity (cm/s) | 3.8 | 8 | 13 | 33 |
| Intensity | VI | VII | VIII | IX |

Table 3. Conversions to macroseismic intensity (MCS) of different peak values for different magnitudes [I(MSK)= ~ (5/6) I(MCS)].

Parametric test with Strike-receiver angle = 45°

| Magnitude | 5.5 | 6 | 6.5 | 7 |
|---|---|---|---|---|
| Peak Velocity (cm/s) | 5 | 8 | 22 | 45 |
| Intensity | VII | VIII | IX | IX |

Table 4. Conversions to macroseismic intensity (MCS) of different peak values for different magnitudes [I(MSK)= ~ (5/6) I(MCS)].

Parametric test with Strike-receiver angle = 0

| Magnitude | 5.5 | 6 | 6.5 | 7 |
|---|---|---|---|---|
| Peak Velocity (cm/s) | 5 | 12 | 25 | 58 |
| Intensity | VII | VIII | IX | >X |

Table 5. Conversions to macroseismic intensity (MCS) of different peak values for different magnitudes [I(MSK)= ~ (5/6) I(MCS)].

The figures, 7a,b,c, briefly illustrate the operating principles of MoSE gates. In case of an extraordinary tide event the gates, normally hidden at the sea-bottom (fig. 7a), are emptied from the water with compressed air (black arrows in fig. 7b), this process lasts about 30 minutes while the revers process, to reopen the barriers, requires about 15 minutes. Then the gates are standing up for about 4-5 hours during the high tide event and, as shown in fig. 7c, they are designed to stay in oblique position (bended towards the lagoon) sustained mostly by the Archimede's effect

Considering these design features we can ask again which would be the consequences of a tsunami wave hitting the gates from the lagoon side, both in the case of a first arriving peak and of a first arriving trough.

Lagoon                              Open sea

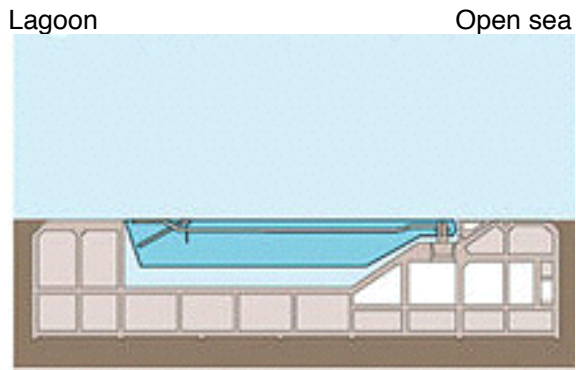

Fig. 7a. MoSE gate during normal tide

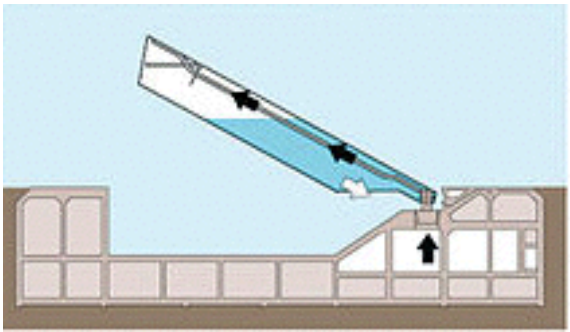

Fig. 7b. MoSE gate lifting up before an extraordinary high tide event

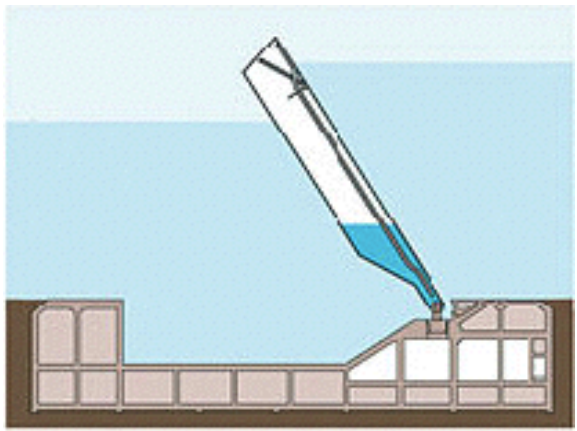

Fig. 7c. MoSE gate standing up during an extraordinary high tide event